\documentclass[12pt,preprint]{aastex}

\shorttitle{Helicity Condensation in the Corona and Wind}
\shortauthors{Antiochos}

\begin{document}

\title{Helicity Condensation as the Origin of Coronal and Solar Wind Structure}


\author{S. K. Antiochos}
\affil{NASA Goddard Space Flight Center, Greenbelt, MD, 20771}
\email{spiro.antiochos@nasa.gov}


\begin{abstract}

  Three of the most important and most puzzling features of the Sun's
  atmosphere are the smoothness of the closed field corona, the
  accumulation of magnetic shear at photospheric polarity inversion
  lines (PIL), and the complexity of the slow wind. We propose that a
  single process, helicity condensation, is the physical mechanism
  giving rise to all three features. A simplified model is presented
  for how helicity is injected and transported in the closed corona by
  magnetic reconnection. With this model we demonstrate that helicity
  must condense onto PILs and coronal hole boundaries, and estimate
  the rate of helicity accumulation at PILs and the loss to the
  wind. Our results can account for many of the observed properties of
  the closed corona and wind.

\end{abstract}




\section{Introduction}

A classic, but puzzling feature of the Sun's high-temperature ($>
1$MK) atmosphere is its apparent lack of complexity. High-resolution
XUV and X-ray images of the closed-field corona, such as those from
the Transition Region and Coronal Explorer (TRACE) mission, invariably
show a smooth collection of loops \citep[e.g.][]{schrijver99}. If the
underlying photospheric flux distribution is highly structured with
several polarity regions, then the topology of the loops in the corona
will appear complex in XUV images, but this is only due to seeing
through multiple flux systems. The surprising result is that the loops
in any one flux system, such as a bipolar active region, are generally
not observed to be twisted or tangled.  Since the coronal field is
line-tied to the photosphere, which is undergoing continuous chaotic
motions due to the convective flows, it would seem that geometrical
complexity should eventually appear in the coronal field. Instead, the
magnetic field appears to remain laminar and not far from a potential
state over most of the corona. Extrapolations of the field do indicate
the presence of coronal currents \citep{leka96,tian08}, but
these are large-scale volumetric currents that produce only a global
shear or twist rather than field line tangling
\citep{schrijver07}.

The observation that loops are untangled, at least on present
observable scales, is especially surprising given that the corona is
being heated continuously. The standard theory for the heating is that
the energy is due to stressing of the coronal field by the random
motions of the photospheric footpoints. This is the basic idea of
Parker's nanoflare model and similar theories
\citep{parker72,parker83,parker88,vanball86,mikic89,berger91,rappazzo08}.
The photospheric motions are postulated to tangle and braid the field
lines, producing small scale current sheets, which then release their
energy via reconnection. A great deal of work has been done applying
this nanoflare scenario to coronal observations with considerable
success \citep{klimchuk06}.

The problem with such reconnection-heating models is that any helicity
injected into the corona as a result of the motions is expected to
survive, because reconnection in a high Lundquist-number system like the
corona conserves magnetic helicity
\citep{taylor74,berger84}. Consequently, even if it is
injected on scales below present-day resolution, $ < 1$ arcsecond, the
helicity should build up and appear as twisting or tangling of the
large-scale coronal field. Note also that even if the degree of
tangling required for the heating is small -- for example, Parker
estimates a misalignment angle between the reconnecting stressed field
and the initial potential state of only $20^\circ$ or so
\citep{parker83} -- any net helicity injected by the stressing
should continue to accumulate and eventually produce large-scale
observable effects. We conclude, therefore, that both the basic
observations of photospheric motions and the reconnection theories for
coronal heating imply that the coronal field should have a complex
geometry, in direct disagreement with observations.

There are, at least, two seemingly likely explanations for this
disagreement.  The first is that photospheric motions produce equal
and opposite helicity everywhere, so that no net helicity is injected
into the corona. Nanoflare reconnection then simply cancels out the
positive and negative helicities.  This explanation, however,
has both observational and theoretical difficulties.  Numerous
observations imply that photospheric motions, including flux
emergence, do inject a net helicity into each hemisphere. For example,
observations of prominence structure
\citep{martin92,rust94,zirker97,pevtsov03} and of active region vector
fields \citep{seehafer90,pevtsov95} indicate a strongly preferred sign
for the helicity injected into each solar hemisphere, possibly related
to the differential rotation \citep{devore00}.  Furthermore, theory
and numerical simulations \citep{linton01} have shown that, for
magnetic flux tubes with parallel axial fields, reconnection occurs
only if the tubes have the same sign of helicity, the so-called
co-helicity case as discussed by \citet{yamada90}. Flux tubes with
opposite helicity only bounce when they collide \citep{linton01};
therefore, reconnection cannot cancel out positive and negative
injected helicity in interacting coronal loops. This point will be
clarified in Figure 2 below.

The other possible explanation for the lack of helicity buildup is
that the heating is due not to reconnection, but to true diffusion, in
which case helicity is not conserved. This hypothesis was proposed by
\citet{schrijver07} to explain the TRACE images. He argued that
continual reconnection induced by the rapidly varying field of the
magnetic carpet \citep{harvey85,schrijver97} causes the chromosphere and
transition region to act like a high-resistivity layer.  Coronal loop
field lines can slip along this layer and, thereby, lose their
tangles.  This explanation, however, also has theoretical and
observational difficulties. Reconnection is not physically equivalent
to diffusion, no matter how frequent the reconnection. Diffusion does
destroy helicity and will relax a magnetic field back down to its
minimum energy potential state, but reconnection can relax the system
down only to some state compatible with total helicity conservation,
such as a linear force-free state \citep{taylor74,taylor86}. In fact,
for the line-tied corona, we have argued that helicity imposes very
stringent constraints on the possible end state of a system undergoing
reconnection relaxation \citep{ska02}.

Furthermore, observations imply that whenever helicity is, indeed,
present in the corona, it does not show evidence for diffusive
decay. The largest concentration of coronal helicity is in filaments
and prominences, or more precisely, in the strongly sheared field that
defines a filament channel \citep[e.g.][]{tandberg95,mackay10}.
Although there is still debate over the exact topology of the filament
channel magnetic field; in particular, whether it is a sheared arcade
\citep{ska94} or a twisted flux rope, the models agree that for
a physically realistic 3D topology all the filament flux must be
connected to the photosphere.  Consequently, if the chromosphere or
transition region really did contain a high-resistivity layer, the
filament channel shear would simply disappear by field-line
slippage. Such slippage is never observed; if anything, filament shear
seems to increase continuously until it is ejected from the corona with
a filament eruption/CME.
 
From the discussion above, we conclude that a net helicity is injected
into each coronal hemisphere by the photosphere, and that reconnection
preserves this helicity. But, in that case, where does the helicity
go? In a sense, the answer is obvious: The helicity injected into the
closed-field corona must end up as the magnetic shear in filament
channels.  These are the only locations in the corona where the
magnetic field is strongly non-potential and, hence, has a strong
helicity concentration.

Although this answer is intuitively appealing, it seems extremely
unlikely. It naturally raises the long-standing questions: What
exactly are filament channels and how do they form? Along with laminar
coronal loops, filament channels are also classic but puzzling
features of the Sun's atmosphere. These structures consist of low-lying
magnetic flux centered about photospheric polarity inversion lines
(PIL), in which the chromospheric and coronal magnetic field lines run
almost parallel to the inversion line rather than perpendicular, as
expected for a potential field \citep{rust67,leroy83,martin98}. Direct
measurements of the filament vector field both in the photosphere
\citep{kuckein12} and corona \citep{casini03} show that the component
parallel to the PIL is dominant. The channels have narrow widths, of
order 10 Mm, but their lengths can be greater than a solar diameter for
PILs that encircle the Sun.  Filament channels are very common,
invariably appearing about any long-lived PIL, both in active regions
and quiet Sun. It should be emphasized that the channels are much more
common than observable filaments and prominences, which require the
presence of substantial amounts of cold plasma as well as the magnetic
shear.

Two general mechanisms have been proposed for filament channel
formation. One mechanism is flux emergence, specifically the emergence
of a sub-photospheric twisted flux rope. Most simulations of flux rope
emergence find that the resulting structure in the corona is a sheared
arcade localized near the PIL
\citep[e.g.][]{manchester01,fan01,magara03,archontis04,leake10,fang12}.
The basic process is straightforward; the twist component of the
sub-photospheric flux rope emerges to become the overlying
quasi-potential arcade in the corona, while the axial sub-surface
component emerges to become the shear field of the filament
channel. It is interesting to note that, in general, the resulting
filament channel in the corona is {\bf not} a twisted flux rope, but a
sheared arcade, because the concave-up portion of the flux-rope field
lines stays trapped below the surface even in 3D \citep{fang12}.

Although flux emergence can yield a sheared arcade, there are major
theoretical and observational difficulties with this process as the
general mechanism for filament channel formation.  First, the
simulations have yet to show that flux emergence agrees quantitatively
with the amount of flux and degree of shear measured in observed
filament channels. In fact, \citet{leake10} argue that only a small
amount of axial flux emerges, at least in 2.5D simulations, far too
small to account for the magnetic free energy in observed filament
channels. Similar conclusions have been reached from recent 3D
simulations, as well \citep{fang12}. A much greater problem for the
model is that filament channels are frequently observed to form in
regions where there is no apparent flux emergence, such as at PILs
between decaying regions and high latitude PILs
\citep[e.g.][]{mackay10}. Moreover, the emergence of a simple
bipolar active region rarely produces a filament channel at its
PIL. The channel usually forms only well after the end of the flux
emergence, when the active region has decayed and dispersed to
interact with surrounding flux regions. Consequently, flux emergence
cannot be the only mechanism for filament channel formation.

The second, and perhaps, the most popular mechanism that has been
proposed for filament channel formation is flux cancellation
\citep{martin98}. The basic picture is that large-scale shear due to
differential rotation or flux emergence concentrates at PILs as
opposite-polarity photospheric flux converges and cancels there. Note
that this mechanism inherently requires reconnection at the
photospheric PIL in order to form low-lying loops that can sink and
disappear and concave-up loops that can rise into the corona
\citep{vanball89}.  A fundamental difficulty with such reconnection,
however, is that it produces a twisted flux rope in the corona just
like flare reconnection produces the highly twisted flux rope of a
CME. On the other hand, high resolution observations of filaments both
from the ground \citep{lin05} and space \citep{vourlidas10} show a
field geometry consisting of long, parallel strands, with no evidence
of twist or tangling. In fact, empirical models for filaments derived
solely from observations, have a laminar field geometry {\it exactly
  like that of the TRACE loops}, except that the field lines are
stretched out and primarily horizontal rather than arched (Martin
ref). It has been suggested that the large twist component resulting
from reconnection may diffuse away \citep{vanball04}, but as argued
above, any diffusion would also decrease the shear component, contrary
to observations. Note also that the twist produced by flare
reconnection, which is physically identical to flux cancellation
reconnection, is never observed to diffuse away, but is measured to
persist out to 1 AU \citep{kumar96,qiu07}.

In addition to the lack of observed twist, the prevalence of filament
channels poses severe difficulties for the flux cancellation model
and, indeed, for any model. As stated above, filament channels are
ubiquitous, appearing over all types of PILs ranging from the most
complex and strongest active regions to very quiet high-latitude
regions. In fact, it is not uncommon to observe a filament channel
that continues unbroken over a PIL that passes from an active region
into neighboring quiet region with the cold material, itself,
transitioning smoothly from a typical low-lying active region filament
to a high-lying quiet sun filament \citep[e.g.][]{su12}. Given these
observations, it seems improbable that filament channels are due to
some phenomenon in the plasma-dominated photosphere, because the
dynamics there are insensitive to the magnetic field structure and, in
particular, to whether a PIL is present or not. This is especially
true in the weak-field regions where quiescent prominences typically
form \citep{klimchuk87}. It seems much more likely that filament
channel formation is due to some generic process occurring in the
magnetically-dominated corona and upper chromosphere.

We propose that the origin of filament channels is the
reconnection-driven evolution of helicity injected into the
closed-field corona. This hypothesis seems counterintuitive,
because coronal-loop helicity is injected on small scales
more-or-less uniformly throughout the corona, whereas filament
channels are coherent structures, localized only around PILs
and extending to very large scale. In this paper, we describe a
process, helicity condensation, that performs exactly the required
transformation of small-scale coronal loop helicity into large-scale
filament-channel shear.  Helicity condensation keeps coronal loops
laminar while shearing filament channels. 

Furthermore, helicity condensation may be responsible for much of the
complex structure and dynamics observed in the slow solar wind. We
argued above that the helicity injected into the corona must end up in
filament channels, but in regions containing coronal holes,
another possibility is that some helicity is ejected out into the wind by
the opening of closed flux at the coronal hole boundary. Such helicity
transfer is implicitly present in the S-Web model for the slow wind
\citep{ska11,ska12}, which postulates that this wind is due to
continual dynamics of the open-closed flux boundary. If so, then
helicity condensation also will play a major role in the origin and
properties of the slow solar wind.

We describe below the basic process of helicity condensation and
derive estimates of its effectiveness in the Sun's corona.

\section{A Model for Helicity Injection and Transport}

In order to understand how magnetic helicity is likely to evolve in
the corona, we must first consider the injection process.  Assume, for
the moment, that the photospheric flux distribution consists of only
two polarity regions as shown in Figure 1: a negative northern
hemisphere and a positive south, so that all the flux closes across
the equatorial PIL (dashed line). The yellow arches in the figure
denote two arbitrary small flux tubes corresponding to coronal loops
or to the strands inside an observable coronal loop.  The quasi-random
photospheric motions will introduce small-scale structure and inject
helicity to this coronal field. Helicity will also be injected by
large-scale motions, such as differential rotation, and by flux
emergence/cancellation, such as the magnetic carpet, but for
simplicity let us model the injection as due to the continual
small-scale photospheric motions, in particular, the granular or
supergranular flows. Note that if the magnetic carpet dynamics do not
change the net coronal flux, their effect on the coronal helicity can
be captured by effective photospheric motions. Furthermore, recent
analysis of high-resolution Solar Dynamics Observatory (SDO) data
indicates that the bulk of the helicity injected into active regions is
due to photospheric motions rather than flux emergence \citep{liu12}.

Following the arguments of \citet{sturrock81}, the energy and,
certainly, the helicity injected into coronal loops by stochastic
horizontal flows at the photosphere will be primarily in the form of
twist. Therefore, we model the motions as a set of randomly located
and randomly occurring rotations that have fixed spatial and temporal
scales. The true photospheric motions are more complex than a set of
fixed-scale rotations, but we are interested only in that part of the
flow that injects helicity to the corona. Note also that there is some
evidence for exactly the pattern of photospheric rotations of Fig. 1
in measurements of the vorticity of the supergranulation
\citep{duvall00,gizon03,komm07}.

It has been known since the time of \citet{hale27} that sunspot whirls
have a clear hemispheric preference, counterclockwise in the north and
clockwise in the south \citep{pevtsov95}, indicating a
preferred sense for the helicity of the subsurface solar motions. The
same helicity preference, negative in the north and positive in the
south, has been well documented to occur in all types of coronal
magnetic structures ranging from quiet Sun field to active region
complexes \citep{pevtsov03b} and has been observed out in the
heliospheric magnetic field \citep{bieber87}. As shown in Figure 1, this
hemispheric ``rule'' is in the same sense as would be expected from
the surface differential rotation, but the actual mechanism is still
not clear.  In any case, we expect there to be a preferred sense
to the helicity injecting rotations as illustrated in Figure 1. Note
that this is only a preference; a fraction of the rotation in each
hemisphere could well have the ``unpreferred'' sense.

The motions shown in Figure 1 have a number of interesting
implications for the coronal field. Assuming, for simplicity, that the
rotations are solid body, have size $d$, and have magnitude,
$\Theta$, then each rotation of a photospheric flux tube with axial
flux, 
\begin{equation}
\Phi_d = \pi d^2 B_p / 4, 
\end{equation}
produces in the corona a twist flux, 
\begin{equation}
\Phi_t = \Theta \Phi_d / \pi,
\end{equation}
where $B_p$ is the average normal field at the photosphere. In open
field regions (not shown in the Figure), this twist flux simply
propagates outward, resulting in a net helicity to the turbulence in
the fast wind \citep[e.g.][]{leamon98}. We will discuss the
implications for the slow wind below. In the closed field
regions, however, the coronal loops acquire a twist component to their
magnetic field, as shown in the Figure. If the loop is perfectly
symmetric about the equator, then on average, the twists imposed by
the two footpoints cancel out so that no net helicity is
injected. Basically, the loop is twisted at one end, but untwisted at
the other. On the other hand, if the loop has both footpoints in one
hemisphere, as is usually the case when the PIL is not exactly at the
equator, then the twist from each footpoint will add. Even if the loop
is transequatorial, we do not expect any symmetry for a real coronal
loop, so a net twist will still be produced by the footpoint
motions. Note also, that the effect of any unpreferred-sense
rotations (clockwise in the north and counterclockwise in the south)
is only to decrease the rate of twisting.  To first order, the
unpreferred rotations simply untwist the loops, but it should be
emphasized that since the rotations are time varying, they can create
higher order topological structure in the field even if the net
injected helicity vanishes. All higher order topological features,
however, such as the braiding of three flux tubes, are not conserved
by reconnection \citep[e.g.][]{pontin11} and are not expected to build
up in the corona.

Consider now the interaction of the two twisted flux tubes of Fig. 1
due to some random motion that causes them to collide. In fact, the
twist itself will cause the flux tubes to expand and interact. Since
their main axial fields are parallel, only the twist components of the
flux tubes can reconnect.  If the tubes have the same sense of twist
(helicity), then at the contact point between the tubes, their twist
components will be oppositely directed and, hence, will
reconnect. This is illustrated in the three sequences of Fig. 2, where
the red and blue circles correspond to field lines of the twist
magnetic component. In the top sequence the twist components are in
the same sense, so they are oppositely directed at their contact
point; consequently, they will reconnect there. Another way of
understanding this result is to note that for interacting tubes with
the same helicity, the photospheric rotations have a stagnation point
between them. It is well known that such stagnation point flows lead
to exponentially growing separation of magnetic footpoints and, hence,
to exponentially growing currents in the corona \citep[e.g.][]{ska97},
which can drive efficient reconnection.  The effect of this reconnection
is to spread the twist component over the flux of the two tubes, in
other words, the two tubes merge into one globally twisted tube as
illustrated in the Figure and as found in simulations of flux-tube
collisions \citep{linton01}.

On the other hand, if the tubes have opposite twist (helicity), then
at their point of interaction the twist components are parallel. As
illustrated in the second sequence of Figure 2, there is no
reconnection in this case. The tubes simply ``bounce''
\citep{linton01}. This result emphasizes the point made in
\citet{ska97} that reconnection in the solar corona is highly
constrained by line-tying. As a result, the coronal magnetic field
cannot simply relax to a minimum energy \citet{taylor74} state, which
for the opposite twist case corresponds to the initial potential
field.

The third sequence in Figure 2 illustrates the effect of continued
interaction of same-helicity flux tubes. If a larger-scale merged flux
tube reconnects with another tube, the result is further merging of
the flux and the spread of the twist to even larger scale. This
reflects the well-known result from turbulence studies that magnetic
helicity tends to cascade upward in scale \citep[e.g.][]{biskamp93}. A
key point is that the scale referred to in our cascade process is the
amount of axial flux, which is closely related to, but not identical
to the spatial scale. As argued directly below, the helicity cascades
up to the largest possible flux scale, which corresponds to all the
axial flux inside a single polarity region.

Let us consider the end result of this reconnection-driven
helicity cascade. Assume a flux system, as in Figure 3, with
simple topology given by a PIL (heavy dashed curve in the Figure) and
a separatrix curve somewhere on the photosphere (light dashed curve)
that defines all the flux that closes across the PIL. There must be
additional PILs on the photosphere, but these are not shown. If the
separatrix curve lies in the north, then the flux system is entirely
in the north and the twist injected by the photosphere will be
predominately counter-clockwise. Since only the relative footpoint
motions are significant, we can assume without loss of generality that
all the twist is imparted inside the PIL as shown in the Figure.

The expected evolution of this twist is seen in Figure 4, which shows
a top view of the flux system. As a result of reconnection, the
helicity ``condenses'' onto the largest scale in the flux system, the
PIL, since this encompasses all the flux in the system. The PIL
defines the boundary of the polarity region. We conclude, therefore,
that the net effect of the many small-scale photospheric twists and
the coronal reconnection is to impart a coherent, global twist of the
whole flux system that concentrates at the PIL. This global twist is
not a true physical motion; the large-scale flux system does not
actually rotate as a coherent body, but the photospheric helicity
injection and subsequent transport by reconnection does result in
an effective global rotation of the magnetic field.

The key point is that such a rotation of the whole flux system
corresponds to a coherent localized shear all along the PIL, exactly
what is needed to explain the formation of filament channels. Such an
effective motion produces a channel consisting of field lines that are
sheared but smooth and laminar, with no twist or tangles in agreement
with high-resolution observations of prominence threads
\citep{lin05,vourlidas10}. The coronal reconnection in our helicity
condensation model results in a structure that is the direct opposite
to that of flux cancellation reconnection, which invariably produces a
highly twisted flux rope at the PIL.

Another important point is that the helicity condensation mechanism is
unaffected by the shape of the PIL, in particular whether the PIL
contains so-called switchbacks where it forms a sharp zigzag.  As long
as the flux system defined by the PIL is primarily in one hemisphere,
helicity condensation will form a filament channel with the same
chirality all along that PIL and with roughly the same amount of
shear. This result is in contrast to the predictions of some of the
flux cancellation models \citep{vanball98}, but is in good agreement
with observations \citep{pevtsov03}. Furthermore, since the
photospheric convection in either quiet or active regions is not
observed to change significantly with phase of the solar cycle, the
model predicts that the hemispheric helicity rule should hold
independent of solar cycle. Again, this conclusion appears to be in
good agreement with observations \citep{pevtsov03}, unlike some flux
cancellation models \citep{mackay01}.

\subsection{Rate of Helicity Cascade}

Prominences typically form on time scales of several days, which sets
a constraint that any model must satisfy; therefore, we calculate
below the rate of filament channel formation predicted by the helicity
condensation model. Let us consider a polarity region as in Figure 4
with scale $L$ that is large compared to the helicity injection scale
$d$.  The rate of helicity injection $\eta$ into a flux tube $\Phi_d$
by a photospheric twist of average angular velocity $V_d / d$ is
given by the product of the rate of change of the twist flux of
Eq. (2) and the axial flux $\Phi_d$, (which stays constant):
\begin{equation}
\eta \approx  \Phi_d (V_d / d)  \Phi_d
\end{equation}
\citep[e.g.][]{berger00}. The total helicity injection rate at the
scale $d$ over the whole flux system region with scale $L$ is therefore given by:
\begin{equation}
h_d =  \Phi_d^2 (V_d /d) (L/d)^2 =   (\Phi_L/L)^2 d^2 \, V_d/d,
\end{equation}
where $\Phi_L$ is the flux of the whole system. Following Kolmogorov's
classic theory for hydrodynamic turbulence \citep{kolmogorov41}, we
assume a constant helicity transfer rate at any scale
$\lambda$. Therefore:
\begin{equation}
 h_\lambda = (\Phi_L/L)^2  \lambda^2 V_\lambda / \lambda  = h_d,
\end{equation}
which implies that:
\begin{equation}
V_\lambda =  V_d  (d/\lambda) \ {\rm and, thus}, \   V_L = V_d  (d/L).
\end{equation}

It should be noted that unlike $V_d$, which is the actual velocity of
the photospheric flows, the quantity $V_L$ does not represent a true
plasma velocity. It is only an effective velocity for the transfer of
twist to the largest scale by coronal reconnection. The physical plasma
velocities will likely be dominated by reconnection jets and will have
both larger magnitude and smaller scale than $V_L$. The physical
velocities and kinetic energy are expected to cascade downward, not
upward, in scale. Consequently, the velocity spectrum $V_\lambda$
derived in Eq. 6 may not be directly observable, but the effective
velocity $V_L$ is indeed physically significant. It quantifies the
rate at which magnetic helicity ``condenses'' out of the corona at the
largest scale of the flux system and, hence, $V_L$ corresponds to the
effective shear velocity along the PIL.

Note also, that the helicity cascade process derived above is somewhat
different than the usual hydrodynamic turbulence in which velocity is
injected statistically uniformly at some scale and then cascades down
to where it is dissipated, usually at a kinetic scale. In such an
energy cascade, the injection of kinetic energy at a global scale
results in the slow increase of thermal energy (temperature)
approximately uniformly throughout the system. In our cascade,
however, helicity is injected statistically uniformly at some
intermediate scale and then simply piles up at the largest
global scale. Even though the helicity injection (i.e.,
photospheric motions) are uniform, the cascade produces a localized
spatial structure in the corona. Helicity condensation in the Sun's
atmosphere is a striking example of self-organization in a complex
system.

The time scale for filament channel formation can now be calculated
directly from Eq. (6). We note that for the photospheric helicity
injection the important parameter is the product of the velocity and
the coherence scale of that velocity. Granules typically have $V_d
\sim 1$ km/s and $d \sim 700$ km, while supergranules have: $V_d \sim
.25$ km/sec and $d \sim 30,000$ km. Eq. (6) implies that the product
$V_d \, d$ is the important quantity; therefore, supergranules are
expected to dominate the helicity injection. Taking the whole flux
system to have scale $d/L \sim 10 - 100$ implies that a shear of order
1,000 -- 10,000 km will build up in $\sim 10^5$ s, where we assume
that equal helicity is injected at both ends of a flux tube. These
results indicate that a high-latitude filament channel with typical
shear scales of 100,000 km will form in several days or so, which is
consistent with observations \citep{tandberg95,mackay10}.  Note also
that we expect that the width of the helicity condensation region to
be of order the width of the elemental photospheric rotation, $\sim
15,000$ km for supergranules, which again is consistent with the
observed widths of filaments \citep{tandberg95,mackay10}.

The supergranular rate of helicity injection estimated above is also
consistent with estimates of solar helicity loss to the wind. Taking
the system size $L$ to be of order the solar radius $\sim 10^{11}$ cm,
and the average field strength at the photosphere to be $~ \sim 10$ G,
we derive from Eq. (4) and the numbers above, a helicity injection
rate over the solar surface of $\sim 10^{38}$ Mx/s. Over the course of
a full solar cycle, this yields a total helicity loss of $\sim 3
\times 10^{46}$ Mx, which agrees well with the inferred losses from
observations of CMEs and the wind \citep{devore00}.

\subsection{Implications of the Model}

We conclude from the derivation above that helicity condensation can
account for filament channel formation, at least, in regions that do
not exhibit strong flux emergence. The mechanism can also account for
the observed smoothness of coronal loops. Let $\tau$ be the time scale
required for the system to establish a steady state (except, of
course, at the largest scale $L$ where no steady-state is possible).
We expect that $\tau$ is determined by the twist required to produce
intense current sheets, of order a full rotation or so \citep{ska98},
and not by the rate of reconnection. The driving velocity is only
$\sim$ 0.1\% of the coronal Alfven speed, so that the reconnection
need not be fast in order to keep pace with the driving. For a given
$\tau$, the twist angle produced by the effective velocities of
Eq. (6) scales as:
\begin{equation}
\Theta_\lambda = \tau V_\lambda/\lambda \sim \lambda^{-2}
\end{equation}
This result implies that the corona will exhibit the most structure at
the scale at which the twist is injected (presumably the
supergranular scale), and at the largest scale where the helicity
piles up, the whole length of the PIL. The so-called coronal cells recently
discovered by \citet{sheeley12} appear to be evidence for just this
type of structure separation. These authors observe that the
large-scale corona breaks up into three distinct structures: flux
tubes twisted on a scale of 30,000 km or so, long filment channels
along PILs that typically span the whole Sun, and coronal holes. Our
helicity condensation model is in excellent agreement with these
observations.

An important issue that is raised by the observations and that we have
yet to discuss is the effect on the model of a coronal hole or, more
generally, of an open field region. Note, also, that even if no
coronal hole is present, the simple picture of Figs. 3 and 4 is
topologically incomplete. The PIL cannot be the only boundary that
defines the closed negative polarity region. At the very least, there
must be a point somewhere in the region where a magnetic spine line
connects up to a null point \citet{lau90,ska90,priest96}; thereby,
making this field line effectively open. Of course, real solar
polarity regions tend to have much more complexity often containing
intricate open field areas and corridors \citep{ska11}.

Assume that the polarity region of Figs. 3 contains a coronal hole, as
illustrated in Figure 5. The presence of the coronal hole introduces
subtleties to the calculation of helicity evolution, because the
helicity of a truly open field that extends to infinity is not
physically meaningful. An open field can have arbitrary helicity due
to linkages at infinity where the field vanishes, but the topology
does not. Therefore, let us consider instead a system where all the
field lines remain closed so the helicity is well-defined throughout
the evolution, and let us model the coronal hole as a region where no
photospheric twists are imposed; hence, no helicity is injected into
this region. In an actual coronal hole twist is injected by
photospheric motions, exactly as in closed field regions, but the
twist propagates away at the Alfven speed and presumably has no effect
on the subsequent evolution in the low corona. Consequently, we can
simply model this region as being untwisted.

The analysis of the helicity cascade in the large annular flux region
bounded by the PIL and the coronal hole boundary proceeds exactly as
above. The only difference is that the total twisted area $L^2$ is
replaced by $L^2 - H^2$ where $H$ is the scale of the coronal
hole. Therefore, the largest scale for the helicity is given by:
\begin{equation}
L^\prime = \sqrt{L^2 - H^2}
\end{equation}
and the effective velocity for helicity condensation is given by Eq
(6) above except that $L$ is replaced by $L^\prime$. Also, the twist
spectrum, Eq (7) is unchanged. As long as $H << L$, the presence of
the coronal hole has minimal effect on the filament channel formation
process, and on the smoothness of coronal loops.

It is evident from Fig. 5, however, that helicity condensation does
have an effect on the magnetic field near the coronal hole
boundary. We note that twist, or more accurately magnetic shear, also
condenses at this boundary, but curiously enough the shear has the
sense {\it opposite} to that at the PIL. This result may seem
physically unlikely, but in fact it is mandated by helicity
conservation. The key point is that the shear flux $\Phi_1$ that
condenses onto the PIL encircles {\it all} the photospheric flux in
the polarity region, including that in the coronal hole region
(``CH''). Therefore, the helicity $H_1$ due to this shear flux is
given by:
\begin{equation}
H_1 = \Phi_1 \Phi_{L^\prime} + \Phi_1 \Phi_{CH} ,
\end{equation}
where $\Phi_{L^\prime}$ is the amount of closed photospheric flux
inside the PIL and $\Phi_{CH}$ is the amount of photospheric flux in
the coronal hole, which can be arbitrary compared to
$\Phi_{L^\prime}$. But the CH flux is not twisted and never
contributes to the helicity injection; hence, it should not affect the
helicity condensation at the PIL. The only way to ensure that the CH
flux has no effect is to have a shear flux $\Phi_2$ that condense at
the CH boundary and is exactly equal and oppositely directed to that
at the PIL. Such a shear flux encircles only the CH flux and, thereby,
adds a helicity contribution: 
\begin{equation}
H_2 = -  \Phi_1 \Phi_{CH}, 
\end{equation}
which exactly cancels out that between the PIL shear flux and the
CH. Fig. 5 shows that reconnection would produce just this required
shear flux at the CH boundary. {\it In other words, helicity
  condensation predicts that a filament channel should form at coronal
  hole boundaries, at the same rate as at the PIL but with the
  opposite handedness.}

These results clearly have major implications for observations. At
PILs, the magnetic shear builds up until eventually, it is ejected as
a prominence eruption/CME. We expect that a similar process of buildup
and ejection occurs at the CH boundaries, but with much less explosive
dynamics. The closed field lines near the CH boundary consist of very long,
high-lying loops that form the outer shell of the streamer
belt. Consequently, it requires far less shear and free energy to open
up these loops than to eject the filament channel. We expect that
helicity condensation at CH boundaries results in continual small
bursts of flux opening and closing there, as is required by models for
the slow wind \citep{ska11,ska12}. An interesting prediction is that
the helicity of the closed flux opening at the CH boundary should be
opposite to that of the photospheric injection into the coronal hole
open field lines, i.e., into the fast wind. It may be possible to test
this prediction with in situ measurements.

In summary, we argue that a single deeply-profound process, helicity
condensation, can explain three long-standing observational challenges in
solar/heliospheric physics: the formation of filament channels, the
smoothness of coronal loops, and the origin of the slow
wind. Furthermore, the model implies major new predictions for solar
structure and dynamics. We look forward to many more theoretical and
observational studies of helicity condensation in the Sun's corona and
wind.

\acknowledgments

This work has been supported, in part, by the NASA TR\&T Program. The
work has benefited greatly from the authors' participation in the NASA
TR\&T focused science teams on multiscale coupling and the slow solar
wind. The author thanks C. R. DeVore, J. T. Karpen, and J. A. Klimchuk
for invaluable scientific discussions and J. T. Karpen for help with the
graphics.

\clearpage





\clearpage

\begin{figure}
\epsscale{.95}
\plotone{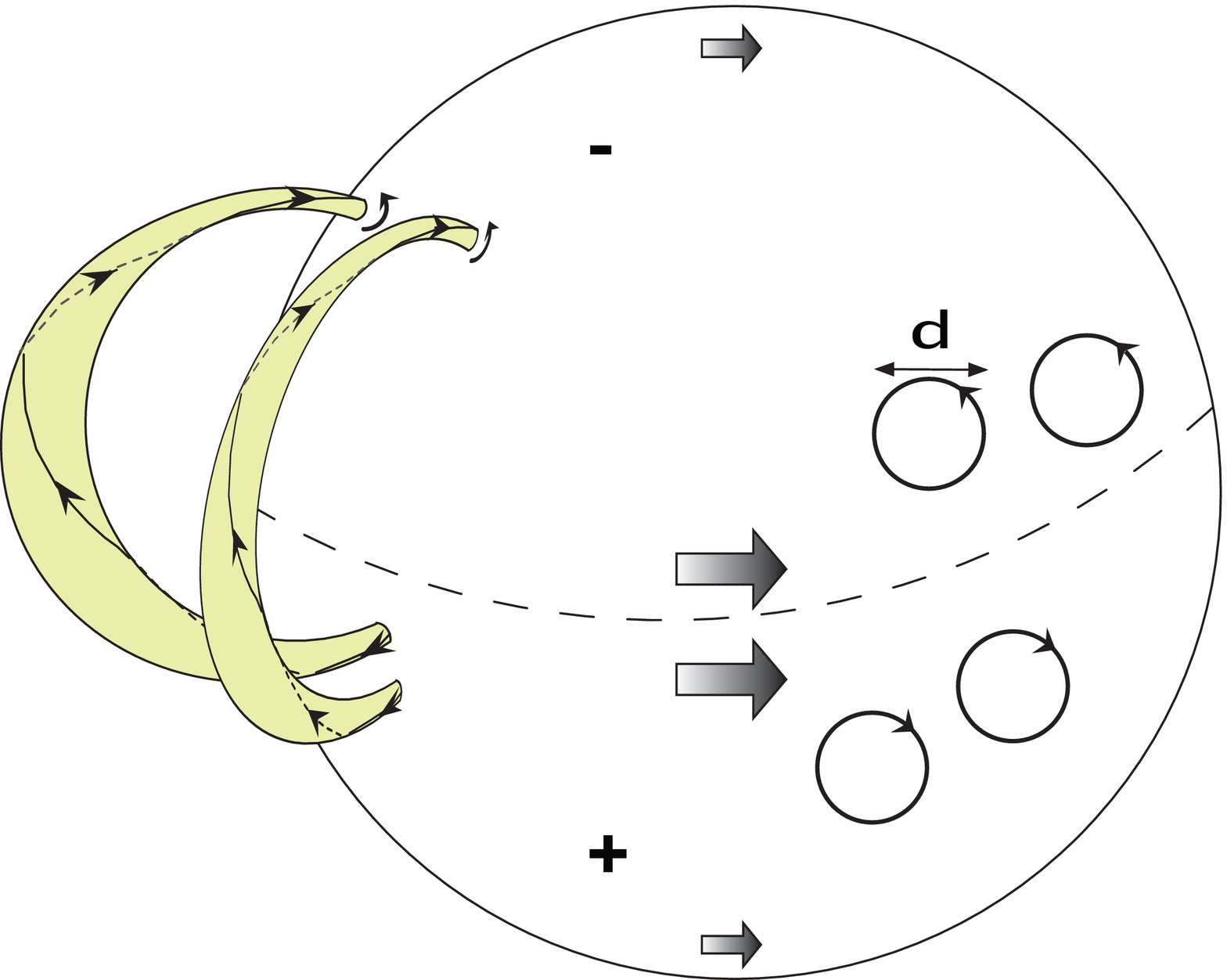}
\caption{ Model for helicity injection into the corona by
  photospheric motions. The primary effect of the motions is to inject
  an effective
  twist of scale {\bf d}. The sense of the twist in each
 hemisphere is determined by the differential rotation, large arrows.
The yellow arches represent two neighboring coronal
  loops (magnetic flux tubes).  
\label{f1}}
\end{figure}
\clearpage

\begin{figure}
\plotone{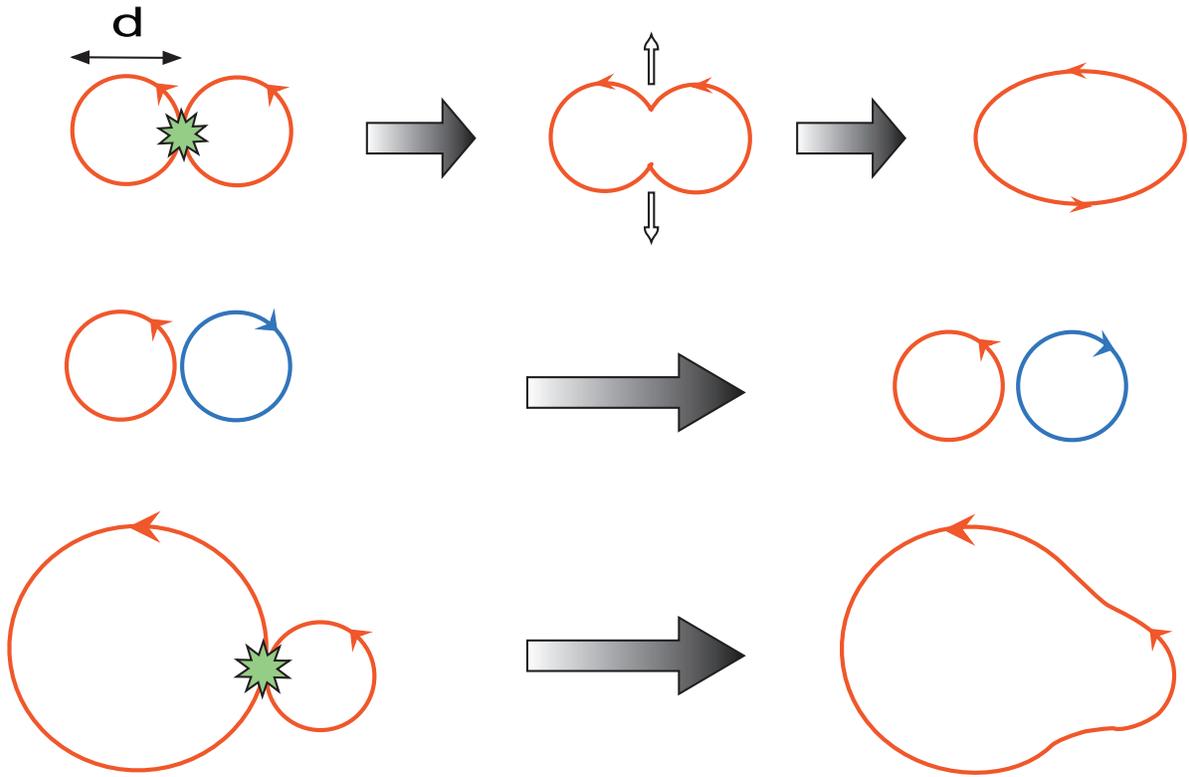}
\caption{ Interaction of the twist component of interacting flux
  tube. Red and blue circles correspond to oppositely-oriented twist
  components of the magnetic field. 
\label{f2}}
\end{figure}
\clearpage

\begin{figure}
\epsscale{.75}
\plotone{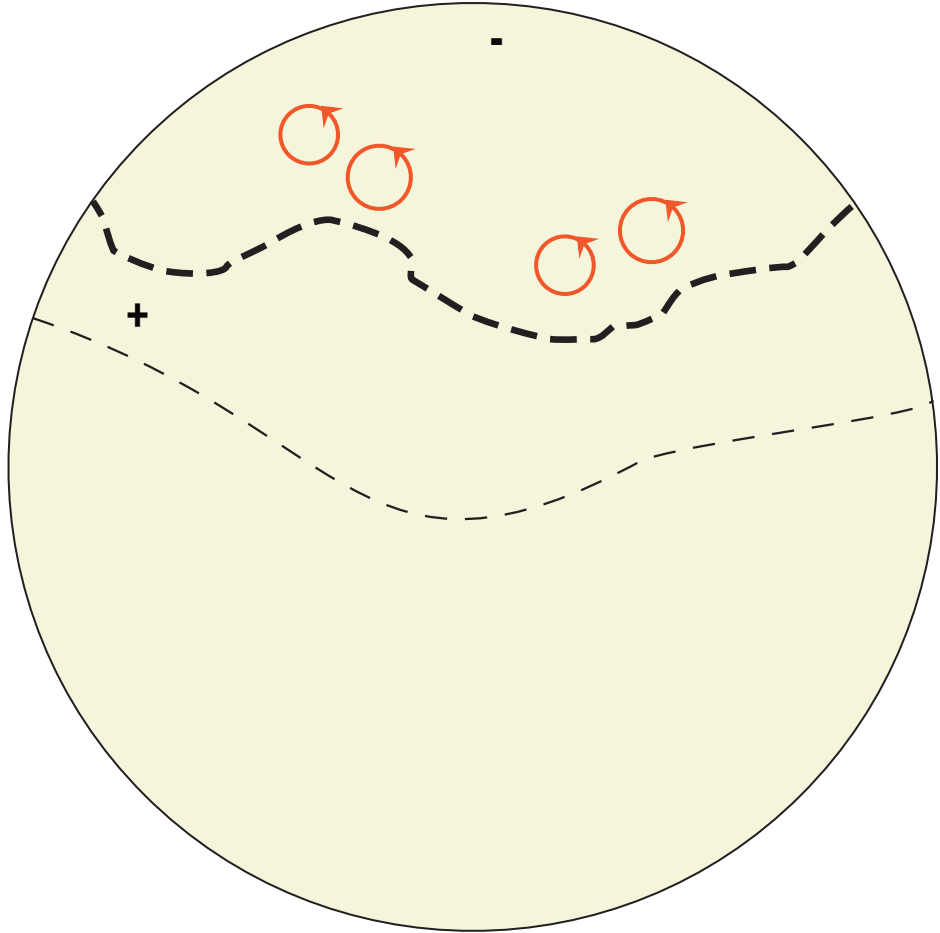}
\caption{ Model of negative polarity flux region fully in the northern
  hemisphere. The dark dashed line correspond to the PIL of the flux
  region.
\label{f3}}
\end{figure}

\begin{figure}
\epsscale{1.0}
\plotone{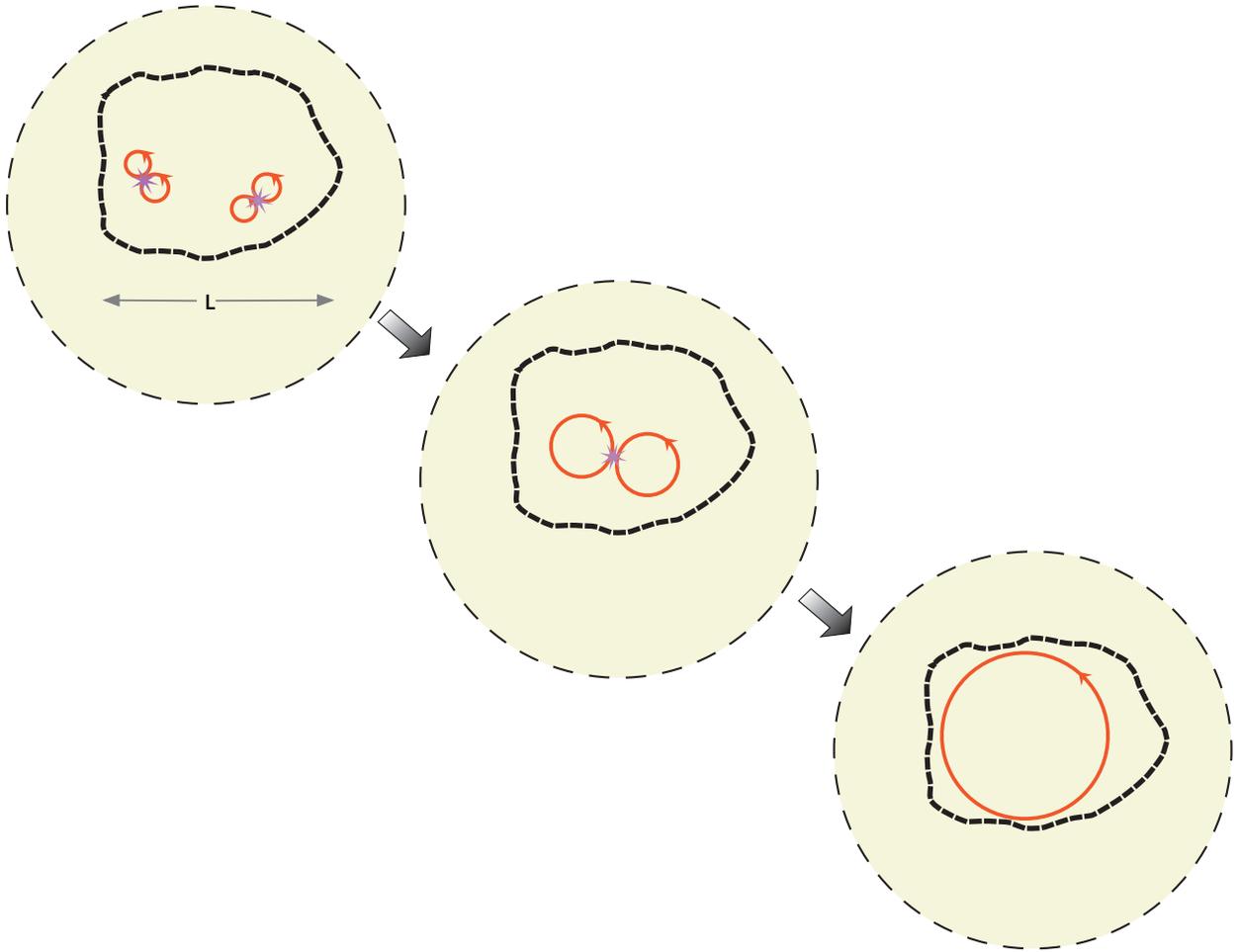}
\caption{ The polarity region of the previous figure as viewed from
  the north pole. 
\label{f4}}
\end{figure}
\clearpage

\begin{figure}
\epsscale{.95}
\plotone{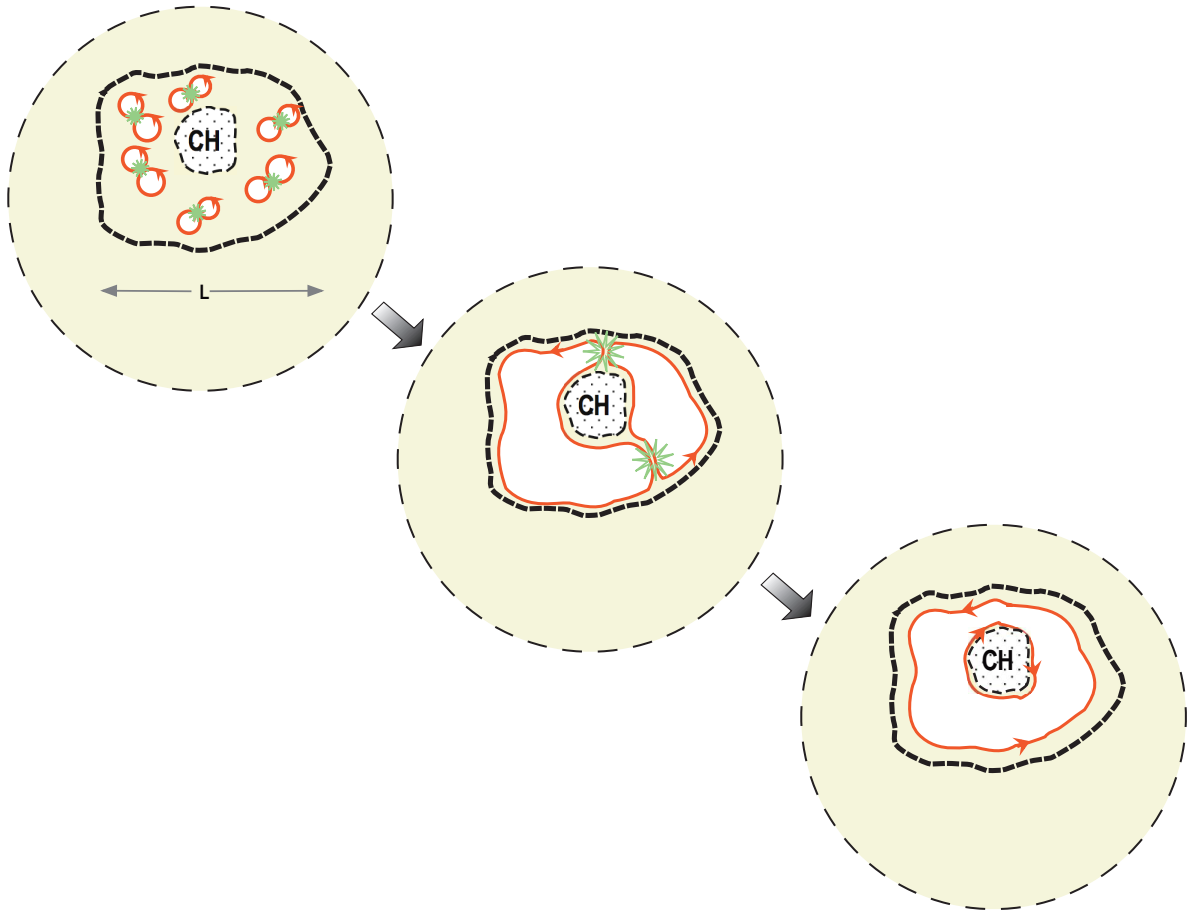}
\caption{ The polarity region of the Figs 3 and 4, but now containing
  a coronal hole region, indicated by ``CH''.
 \label{f5}}
\end{figure}

\end{document}